\documentclass[twocolumn]{aastex62}
\def\ra#1#2#3{#1$^{\rm h}$#2$^{\rm m}$#3$^{\rm s}$}
\def\dec#1#2#3{$#1^\circ#2'#3''$}
\usepackage{natbib}
\usepackage[caption=false]{subfig}

\newcommand{\Msun}{\ensuremath{\textrm{M}_{\odot}}\,}

\graphicspath{{./}{figures/}}

\submitjournal{ApJL}

\shorttitle{He shell double detonation ZTF\,18aaqeasu}
\shortauthors{K. De et al.}
\begin{document}

\title{ZTF\,18aaqeasu (SN\,2018byg): A Massive Helium-shell Double Detonation on a Sub-Chandrasekhar Mass White Dwarf}

\correspondingauthor{Kishalay De}
\email{kde@astro.caltech.edu}

\author[0000-0002-8989-0542]{Kishalay De}
\affil{Cahill Center for Astrophysics, California Institute of Technology, 1200 E. California Blvd. Pasadena, CA 91125, USA.}

\author{Mansi M. Kasliwal}
\affil{Cahill Center for Astrophysics, California Institute of Technology, 1200 E. California Blvd. Pasadena, CA 91125, USA.}

\author{Abigail Polin}
\affil{Department of Astronomy, University of California, Berkeley, CA, 94720-3411, USA.}
\affil{Lawrence Berkeley National Laboratory, Berkeley, California 94720, USA.}

\author{Peter E. Nugent}
\affil{Department of Astronomy, University of California, Berkeley, CA, 94720-3411, USA.}
\affil{Lawrence Berkeley National Laboratory, Berkeley, California 94720, USA.}

\author{Lars Bildsten}
\affil{Department of Physics, University of California, Santa Barbara, CA 93106, USA.}
\affil{Kavli Institute for Theoretical Physics, University of California, Santa Barbara, CA 93106, USA.}

\author{Scott M. Adams}
\affil{Cahill Center for Astrophysics, California Institute of Technology, 1200 E. California Blvd. Pasadena, CA 91125, USA.}

\author{Eric C. Bellm}
\affiliation{DIRAC Institute, Department of Astronomy, University of Washington, 3910 15th Avenue NE, Seattle, WA 98195, USA}

\author{Nadia Blagorodnova}
\affil{Cahill Center for Astrophysics, California Institute of Technology, 1200 E. California Blvd. Pasadena, CA 91125, USA.}

\author{Kevin B. Burdge}
\affil{Cahill Center for Astrophysics, California Institute of Technology, 1200 E. California Blvd. Pasadena, CA 91125, USA.}

\author{Christopher Cannella}
\affil{Cahill Center for Astrophysics, California Institute of Technology, 1200 E. California Blvd. Pasadena, CA 91125, USA.}

\author{S. Bradley Cenko}
\affil{Astrophysics Science Division, NASA Goddard Space Flight Center, Mail Code 661, Greenbelt, MD 20771, USA.}
\affil{Joint Space-Science Institute, University of Maryland, College Park, MD 20742, USA.}

\author{Richard G. Dekany}
\affiliation{Caltech Optical Observatories, California Institute of Technology, Pasadena, CA 91125, USA}

\author{Michael Feeney}
\affiliation{Caltech Optical Observatories, California Institute of Technology, Pasadena, CA 91125, USA}

\author{David Hale}
\affiliation{Caltech Optical Observatories, California Institute of Technology, Pasadena, CA 91125, USA}

\author{U. Christoffer Fremling}
\affil{Cahill Center for Astrophysics, California Institute of Technology, 1200 E. California Blvd. Pasadena, CA 91125, USA.}

\author{Matthew J. Graham}
\affil{Cahill Center for Astrophysics, California Institute of Technology, 1200 E. California Blvd. Pasadena, CA 91125, USA.}

\author{Anna Y. Q. Ho}
\affil{Cahill Center for Astrophysics, California Institute of Technology, 1200 E. California Blvd. Pasadena, CA 91125, USA.}

\author{Jacob E. Jencson}
\affil{Cahill Center for Astrophysics, California Institute of Technology, 1200 E. California Blvd. Pasadena, CA 91125, USA.}

\author{S. R. Kulkarni}
\affil{Cahill Center for Astrophysics, California Institute of Technology, 1200 E. California Blvd. Pasadena, CA 91125, USA.}

\author{Russ R. Laher}
\affiliation{IPAC, California Institute of Technology, 1200 E. California Blvd, Pasadena, CA 91125, USA}

\author{Frank J. Masci}
\affiliation{IPAC, California Institute of Technology, 1200 E. California Blvd, Pasadena, CA 91125, USA}

\author{Adam A. Miller}
\affiliation{Center for Interdisciplinary Exploration and Research in Astrophysics (CIERA) and Department of Physics and Astronomy,
Northwestern University, 2145 Sheridan Road, Evanston, IL 60208, USA}
\affiliation{The Adler Planetarium, Chicago, IL 60605, USA}

\author{Maria T. Patterson}
\affiliation{DIRAC Institute, Department of Astronomy, University of Washington, 3910 15th Avenue NE, Seattle, WA 98195, USA}

\author{Umaa Rebbapragada}
\affiliation{Jet Propulsion Laboratory, California Institute of Technology, Pasadena, CA 91109, USA}

\author{Reed L. Riddle}
\affiliation{Caltech Optical Observatories, California Institute of Technology, Pasadena, CA 91125, USA}

\author{David L. Shupe}
\affiliation{IPAC, California Institute of Technology, 1200 E. California Blvd, Pasadena, CA 91125, USA}

\author{Roger M. Smith}
\affiliation{Caltech Optical Observatories, California Institute of Technology, Pasadena, CA 91125, USA}

\begin{abstract}

The detonation of a helium shell on a white dwarf has been proposed as a possible explosion triggering mechanism for Type Ia supernovae. Here, we report ZTF\,18aaqeasu (SN\,2018byg/ATLAS\,18pqq), a peculiar Type I supernova, consistent with being a helium-shell double-detonation. With a rise time of $\approx 18$\,days from explosion, the transient reached a peak absolute magnitude of $M_R \approx -18.2$\,mag, exhibiting a light curve akin to sub-luminous SN 1991bg-like Type Ia supernovae, albeit with an unusually steep increase in brightness within a week from explosion. Spectra taken near peak light exhibit prominent Si absorption features together with an unusually red color ($g-r \approx 2$\,mag) arising from nearly complete line blanketing of flux blue-wards of 5000\,\AA. This behavior is unlike any previously observed thermonuclear transient. Nebular phase spectra taken at and after $\approx 30$\,days from peak light reveal evidence of a thermonuclear detonation event dominated by Fe-group nucleosynthesis. We show that the peculiar properties of ZTF\,18aaqeasu are consistent with the detonation of a massive ($\approx 0.15$\,\Msun) helium shell on a sub-Chandrasekhar mass ($\approx 0.75$\,\Msun) white dwarf after including mixing of $\approx 0.2$\,\Msun of material in the outer ejecta. These observations provide evidence of a likely rare class of thermonuclear supernovae arising from detonations of massive helium shells.
\end{abstract}


\section{Introduction} 
\label{sec:intro}

In the double-detonation model for Type Ia supernovae (SNe), the explosive detonation of a helium (He) shell on the surface of a sub-Chandrasekhar mass white dwarf (WD) triggers a detonation in the core of the WD, leading to an explosion of the entire star \citep{Nomoto1980,Nomoto1982a,Nomoto1982b,Woosley1986,Woosley1994,Livne1995}. Several key issues in this mechanism have been studied in recent years, including the conditions for the detonation of the He shell (that is accreted from a He-rich companion) and if the detonation in the shell can trigger a detonation in the underlying CO core \citep{Bildsten2007,Fink2007,Fink2010,Sim2010,Shen2014a}. These studies have generally concluded that detonations in the shell are triggered for He shell masses larger than $\sim 0.01$\,\Msun, while at the same time inevitably leading to a detonation of the core \citep{Bildsten2007,Fink2010,Shen2010,Shen2014b}.\\

Consequently, several studies have also explored the observational signatures of these events and if they are consistent with observed diversity of Type Ia SNe \citep{Kromer2010,Sim2010,Woosley2011,Polin2018}. While simulations of the double detonation scenario in bare sub-Chandrasekhar mass CO WDs (i.e. without including the effects of the overlying He shell) have found that these explosions are capable of reproducing the observed diversity of Type Ia SNe \citep{Sim2010,Shen2018}, the results are quite different when including the ashes of the overlying He shell (rich in He burning products) in the radiative transfer calculations \citep{Hoeflich1996,Nugent1997,Kromer2010,Woosley2011,Polin2018}.\\

In particular, these models find that He shell double detonation events exhibit spectra that are strongly influenced by Fe-group line blanketing features from the overlying burned material, thus producing unusually red colors near peak light. These features remain generally inconsistent with the observed variety of Type Ia SNe for the minimum He shell masses that have been previously suggested to detonate the core ($\sim 0.05$\,\Msun; as found in the initial simulations of \citealt{Bildsten2007} and \citealt{Fink2010}). Several solutions to these discrepancies have been proposed, including possible differences in the composition of the burnt He shell (e.g. due to pollution of the initial He shell by C; \citealt{Kromer2010}). Alternatively, it has been suggested that the observed variety of Type Ia SNe could be produced from detonations of thin He shells with even lower masses ($\lesssim 0.01$\,\Msun) that may still detonate the core \citep{Shen2014b, Shen2018, Polin2018}. \\

In this paper, we present observations of ZTF\,18aaqeasu, a peculiar Type I SN that exhibits remarkable similarities to expected signatures of a He-shell double detonation on a white dwarf. Section \ref{sec:observations} presents the observations of the transient. Section \ref{sec:analysis} presents a comparison of this source to known Type Ia SNe. Section \ref{sec:models} presents a comparison of the data to models of He shell detonations presented in \citet{Kromer2010} and a larger grid presented in \citet{Polin2018}. We end with a discussion of the observations in the broader context of Type Ia SNe in Section \ref{sec:discussion}. Calculations in this paper assume a $\Lambda$CDM cosmology with $H_0 = 70$\,km\,s$^{-1}$\,Mpc$^{-1}$, $\Omega_M = 0.27$ and $\Omega_{\Lambda}=0.73$ \citep{Komatsu2011}.

\section{Observations}
\label{sec:observations}
\subsection{Detection and classification}
On 2018 May 04.268\footnote{UT times are used throughout the paper} (MJD 58242.268), ZTF\,18aaqeasu was first detected by the Zwicky Transient Facility (ZTF; \citealt{Bellm2018, Graham2018}) using the 48-inch Samuel Oschin Telescope (P48) at Palomar Observatory in a nightly cadence experiment. This detection was at an $r$-band magnitude of $\approx$ 20.57\,mag and J2000 coordinates $\alpha =$ \ra{12}{23}{21.57}, $\delta =$ \dec{46}{36}{08.3}. The source was not detected on 2018 April 25.198 (MJD 58233.198; 9.07 days before first detection) up to a limiting magnitude of $r \geq 20.11$\,mag.\\

On 2018 May 7, ZTF\,18aaqeasu met machine-learning thresholds and was flagged by a science program filter on the GROWTH Marshal \citep{Kasliwal2018} that is designed to look for transients in the vicinity of nearby galaxies. ZTF\,18aaqeasu was detected in the outskirts of an elliptical galaxy at a redshift of $z = 0.066$ (Figure \ref{fig:photometry}). On 2018 May 8.19, we obtained the first spectrum, which exhibited blue continua with broad absorption features below 5000\,\AA\,(Section \ref{sec:obsSpec}). On 2018 May 14.34, a subsequent spectrum exhibited prominent Si II absorption features as well as a sharp cut-off in flux below 5000\,\AA. As such, due to the strong Si II features and absence of any H features, we tentatively classified the transient as a peculiar Type Ia SN (see \citealt{Filippenko1997} for a review). \\

On 2018 May 22.42, the transient was independently detected at 18.9\,mag by the ATLAS survey \citep{Tonry2018} as ATLAS18pqq. On 2018 May 25, ATLAS reported this event to the Transient Name Server (TNS) and the event was given the IAU name AT\,2018byg. On 2018 Nov 19, we reported the spectroscopic classification and it was re-named as SN\,2018byg. Hereafter, we refer to the source by the name ZTF\,18aaqeasu.\\

\subsection{Optical Photometry}

\begin{figure*}[!ht]
\centering
\includegraphics[width=0.41\textwidth]{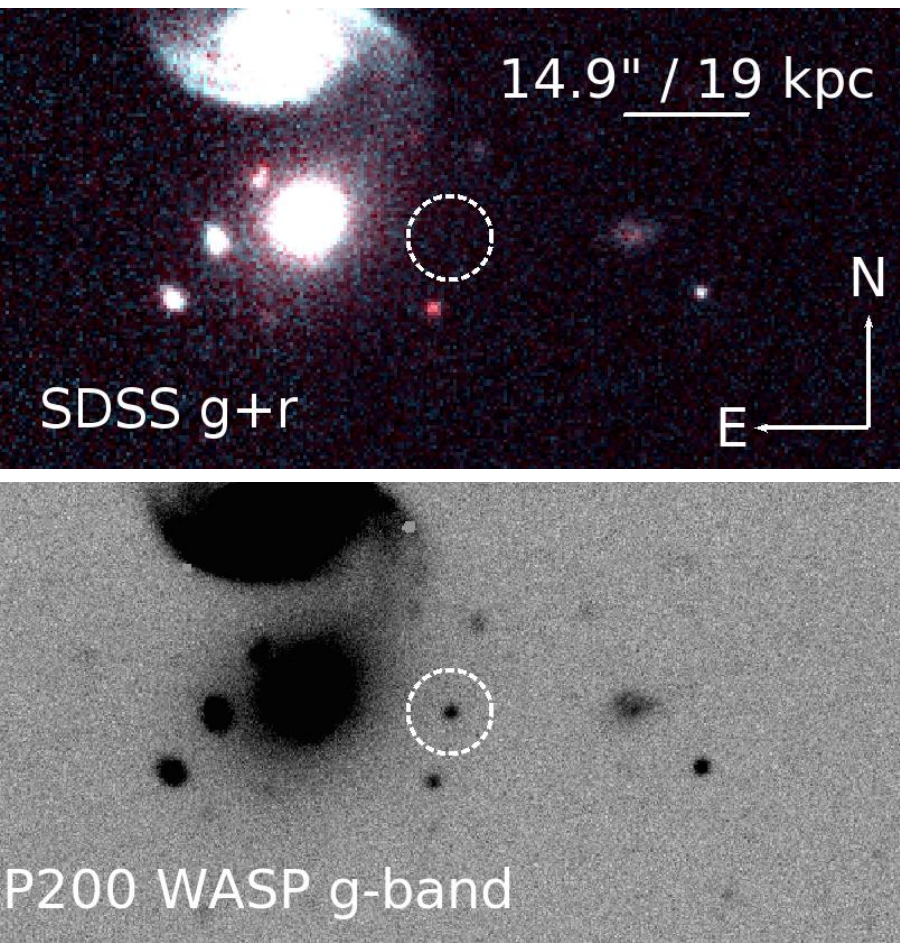}
\includegraphics[width=0.5\textwidth]{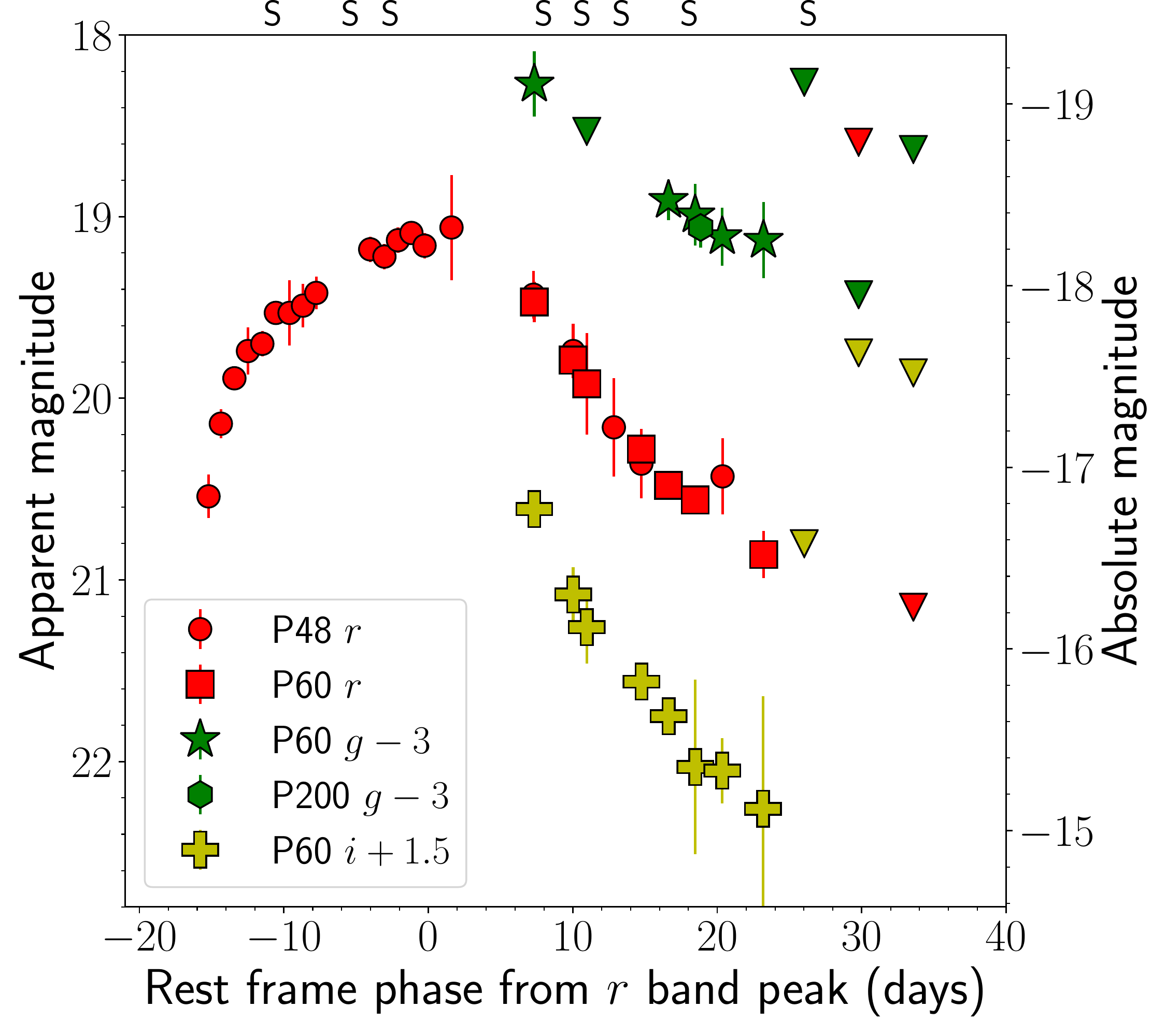}%
\caption{(Left) Detection field and host galaxy of ZTF\,18aaqeasu. The top panel is an archival SDSS image of the region while the lower panel shows an image taken with WASP on P200. The location of the transient is marked with the white circle - it is at a projected offset of $\approx 17.2$\arcsec corresponding to a physical projected distance of $21.9$ kpc at the host galaxy redshift. (Right) Multi-color light curves of ZTF\,18aaqeasu. The inverted triangles are upper limits. The epochs of spectroscopy are marked with `S' on the top axis.}
\label{fig:photometry}
\end{figure*}

We obtained $r$-band photometry of ZTF\,18aaqeasu with the ZTF camera, along with $gri$-band photometry with the Spectral Energy Distribution Machine (SEDM; \citealt{Blagorodnova2018}) mounted on the automated 60-inch telescope (P60; \citealt{Cenko2006}) at Palomar observatory. The P48 images were reduced with the Zwicky Transient Facility Image Differencing pipeline \citep{Masci2018}, which performs host subtracted point spread function (PSF) photometry, while the P60 images were reduced using the pipeline described in \citet{Fremling2016}. We also obtained one epoch of $g$-band imaging with the Wafer Scale Imager for Prime (WASP) instrument mounted on the 200-inch Hale telescope at Palomar observatory on 09 June 2018. These images were reduced with a custom-developed imaging pipeline based in \texttt{python}. \\

We correct all our photometry for galactic extinction for $A_V = 0.032$\,mag from the maps of \citet{Schlafly2011}. We do not correct for any additional host extinction due to the offset location of the transient, and the absence of any Na I D absorption at the host redshift in our spectra. We show the multi-color light curves (magnitudes are in the AB system) of ZTF\,18aaqeasu in Figure \ref{fig:photometry}. For all subsequent discussion, we refer phases with respect to the maximum of the $r$-band light curve.

\subsection{Optical spectroscopy}
\label{sec:obsSpec}
\begin{figure*}
\centering
\includegraphics[width=0.7\textwidth]{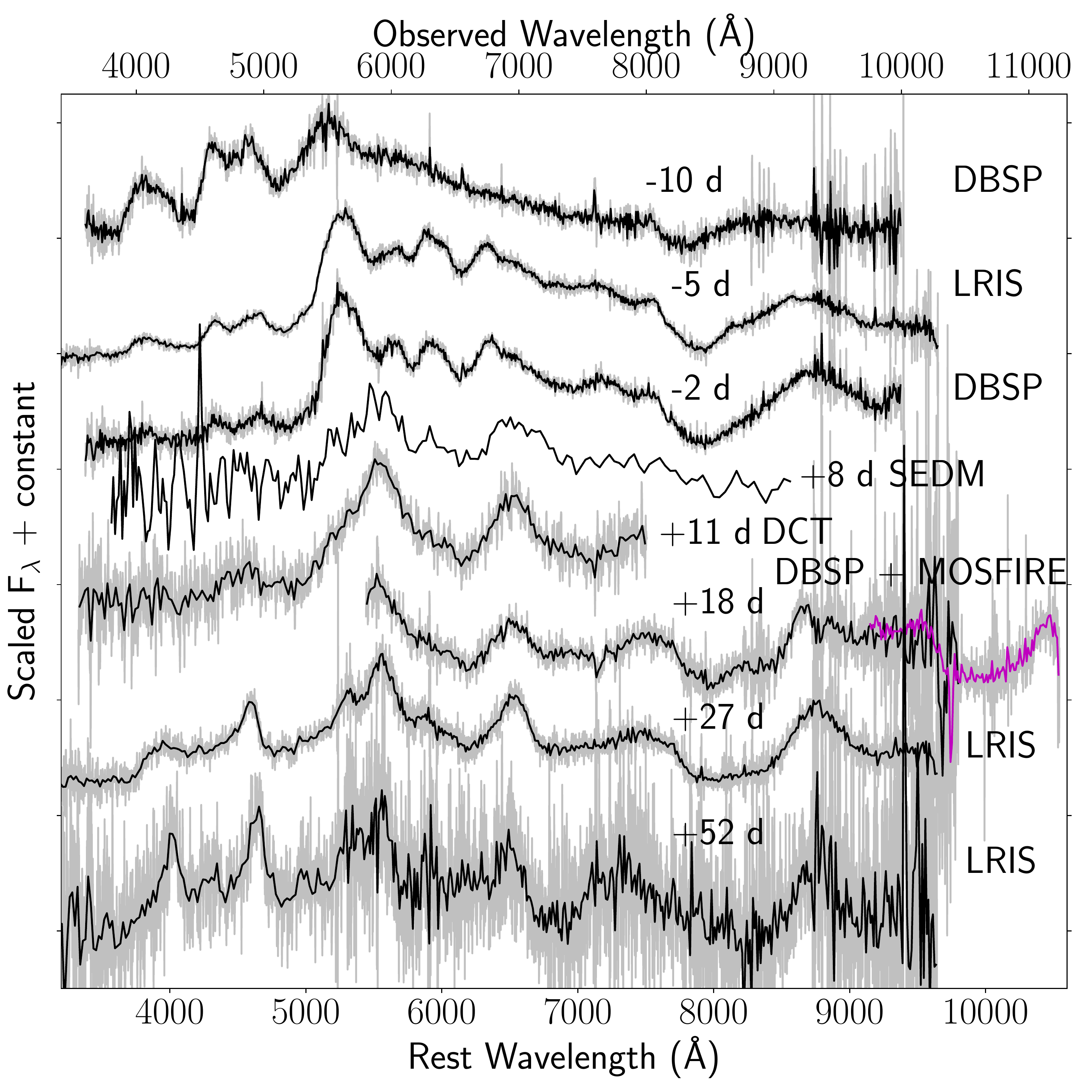}
\caption{Spectroscopic sequence of ZTF\,18aaqeasu. Phases with respect to $r$-band peak and instruments used are indicated next to each spectrum. The black lines are binned from the raw spectra shown in gray lines. The magenta line at $+18$\,days indicates the only NIR spectrum of the source in Y band from MOSFIRE.}
\label{fig:spec_seq}
\end{figure*}

We obtained optical spectroscopic follow-up of the transient starting from $\approx -10$\,d to $\approx +53$\,d after $r$-band peak using the Double Beam Spectrograph (DBSP; on 2018 May 08, 2018 May 17 and 2018 June 08) on the 200-inch Hale telescope \citep{Oke1982}, the Low Resolution Imaging Spectrograph (LRIS; on 2018 May 14, 2018 June 17 and 2018 July 13) on the Keck-I telescope \citep{Oke1995}, the SEDM (on 2018 May 28), and the DeVeny spectrograph (on 2018 May 31) on the Discovery Channel Telescope \citep{Bida2014}. We present our sequence of spectra in Figure \ref{fig:spec_seq}. All spectra will be made publicly available via the WISeREP repository \citep{Yaron2012}.  

\subsection{NIR photometry and spectroscopy}

We obtained Near Infrared (NIR) $JHK$ band imaging of the transient using the Multi-Object Spectrometer for Infrared Exploration (MOSFIRE; \citealt{Mclean2012}) on the Keck-I telescope on 2018 June 03 ($\approx 13$\,days after $r$-band peak). Dithered science exposures of the target field were obtained in each band for a total exposure time of 231\,s, 192\,s and 128\,s in $J$, $H$ and $K_s$ bands respectively. The images were reduced using a custom-built \texttt{python}-based imaging pipeline, and the transient was detected in all three filters. We measure Vega magnitudes of $J = 19.92 \pm 0.10$\,mag, $H = 19.70 \pm 0.25$\,mag and $K_s = 19.29 \pm 0.22$\,mag.  We also obtained a $Y$-band ($9700$--$11100$\,\AA) spectrum of the transient with MOSFIRE on 2018 June 03, for a total integration time of 720\,s. The spectra were reduced with the MOSFIRE Data Reduction Pipeline, and are shown in Figure \ref{fig:spec_seq}.\\

\subsection{Swift XRT observations}

We obtained X-ray follow-up of the transient with the Swift X-ray telescope (XRT; \citealt{Burrows2005}) on the Neil Gehrels Swift Observatory (Swift; \citealt{Gehrels2004}). The Swift observatory observed the location of the transient on 2018 May 27 ($\approx 8$\,days after $r$-band peak) for a total exposure time of $3.8$\,ks. No source was detected at the location of the transient down to a $3\sigma$ limiting flux of $3.5 \times 10^{-3}$\,count\,s$^{-1}$, corresponding to a 0.3--10 keV flux of $1.2 \times 10^{-13}$ ergs cm$^{-2}$ s$^{-1}$. This corresponds to an X-ray luminosity of $\lesssim 1.3 \times 10^{42}$\,ergs\,s$^{-1}$ at the distance of the host galaxy for a photon index of $\Gamma = 2$. In the same observation, no source was detected with the Ultraviolet Optical telescope (UVOT; \citealt{Roming2005}) in the UVW2 filter, down to a 5$\sigma$ limiting AB magnitude of 22.40.

\section{Analysis}
\label{sec:analysis}
\subsection{Photometric properties}
\label{sec:photProp}

\begin{figure*}
\centering
\includegraphics[width=0.49\textwidth]{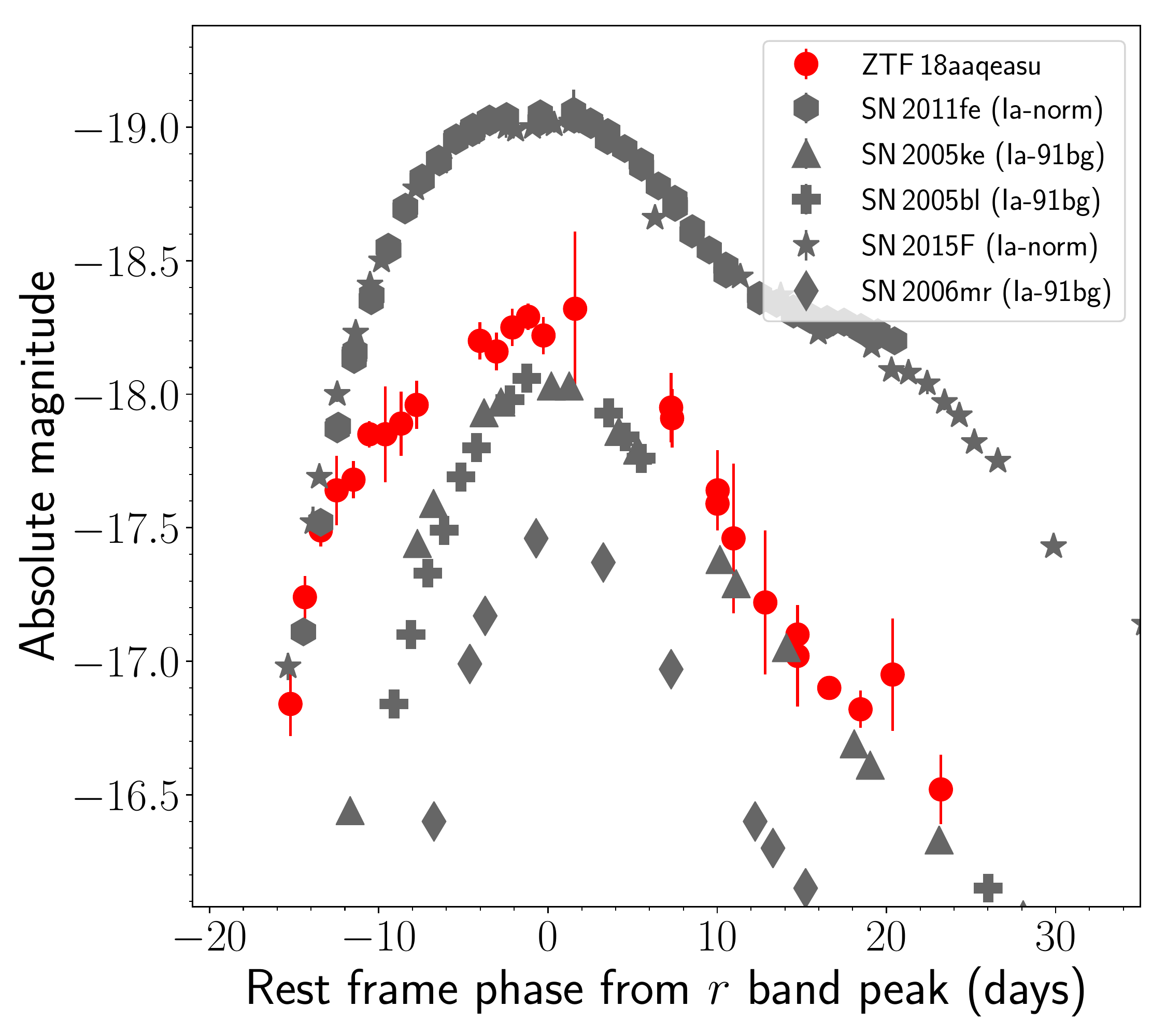}%
\includegraphics[width=0.49\textwidth]{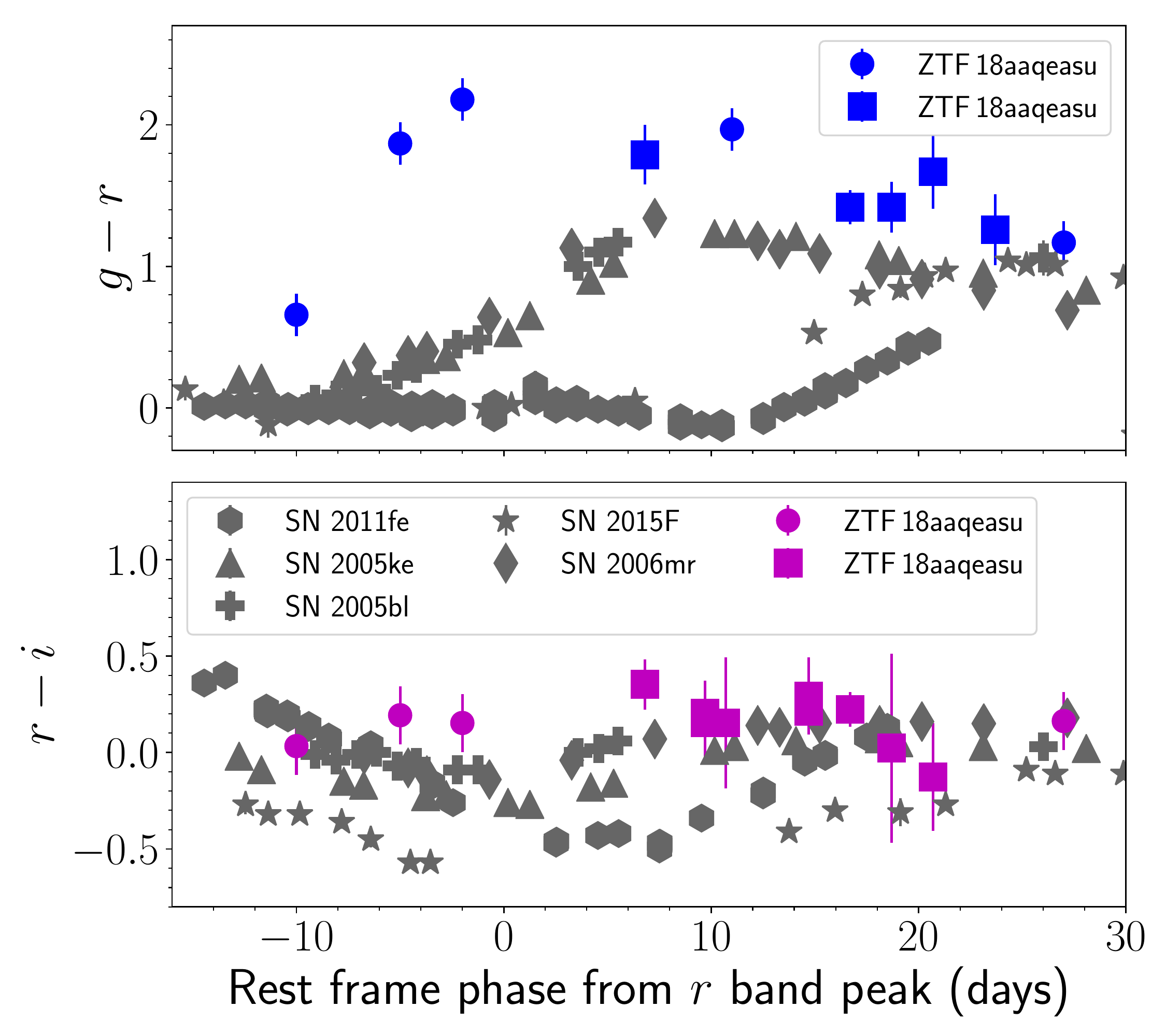}%
\caption{(Left) Comparison of the $r$-band light curve of ZTF\,18aaqeasu to other normal and sub-luminous Type Ia SNe. (Right) Comparison of the $g-r$ and $r-i$ color evolution of ZTF\,18aaqeasu (circles are colors derived from spectra while squares denote colors from photometry) to the same sample of SNe as in the left panel.}
\label{fig:photComp}
\end{figure*}

The transient exhibited a light curve fainter than the normal Type Ia SNe, but similar to the subluminous 1991bg-like events. To this end, we compare the photometric evolution of ZTF\,18aaqeasu to the light curves of the normal Type Ia SNe 2011fe \citep{Nugent2011} and SN 2015F \citep{Cartier2017}, as well as the sub-luminous (SN 1991bg-like) Type Ia SNe 2005ke, 2005bl and 2006mr \citep{Contreras2010}, in Figure \ref{fig:photComp} (data taken from the Open Supernova Catalog; \citealt{Guillochon2017}). The source exhibited an initial fast rise (at $< -10$\,days from peak) of $\approx 0.4$\,mag\,day$^{-1}$ similar to the normal Type Ia SNe but subsequently slowed in its rise transitioning to a sub-luminous SN Ia light curve. \\


By fitting the peak of the light curve with a low order polynomial, we find a best-fit peak time in $r$-band of MJD 58258.49, and a peak absolute magnitude of $M_r = -18.27 \pm 0.04$\,mag. Integrating the total flux in the optical spectrum of the source near peak light (at $\approx -2$\,days; after performing an absolute calibration with respect to $r$-band photometry), we find a lower limit on the peak luminosity of $\approx 2.4 \times 10^{42}$\,ergs\,s$^{-1}$. Using Arnett's law for the synthesized $^{56}$Ni mass \citep{Arnett1985}, we use the peak luminosity to find $^{56}$Ni mass $\gtrsim 0.11$\,\Msun. \\

Although we do not have multi-color photometric coverage before the peak of the $r$-band light curve, we use our well sampled sequence of spectra to construct pre-peak color curves for the transient. We constructed $g-r$ and $r-i$ color curves of the transient by performing synthetic photometry on the spectra, while adding a 10\% uncertainty on the measurements to account for potential inaccuracies in flux calibration. The colors derived from spectroscopy are consistent with contemporaneous multi-color photometry at epochs after peak light. The color evolution is shown in in Figure \ref{fig:photComp}. Comparing the color curves to the other Type Ia SNe, we find that while the $r-i$ color evolution is similar to the subluminous Type Ia SNe in the light curve comparison sample, the $g-r$ color of the transient ($g-r \approx 2$ at peak light) is $\approx 1.5$ mag redder near peak light than all the Type Ia SNe in the comparison sample. 

\subsection{Spectroscopic properties}
\label{sec:specProp}

\begin{figure*}[!ht]
\centering
\includegraphics[width=0.49\textwidth]{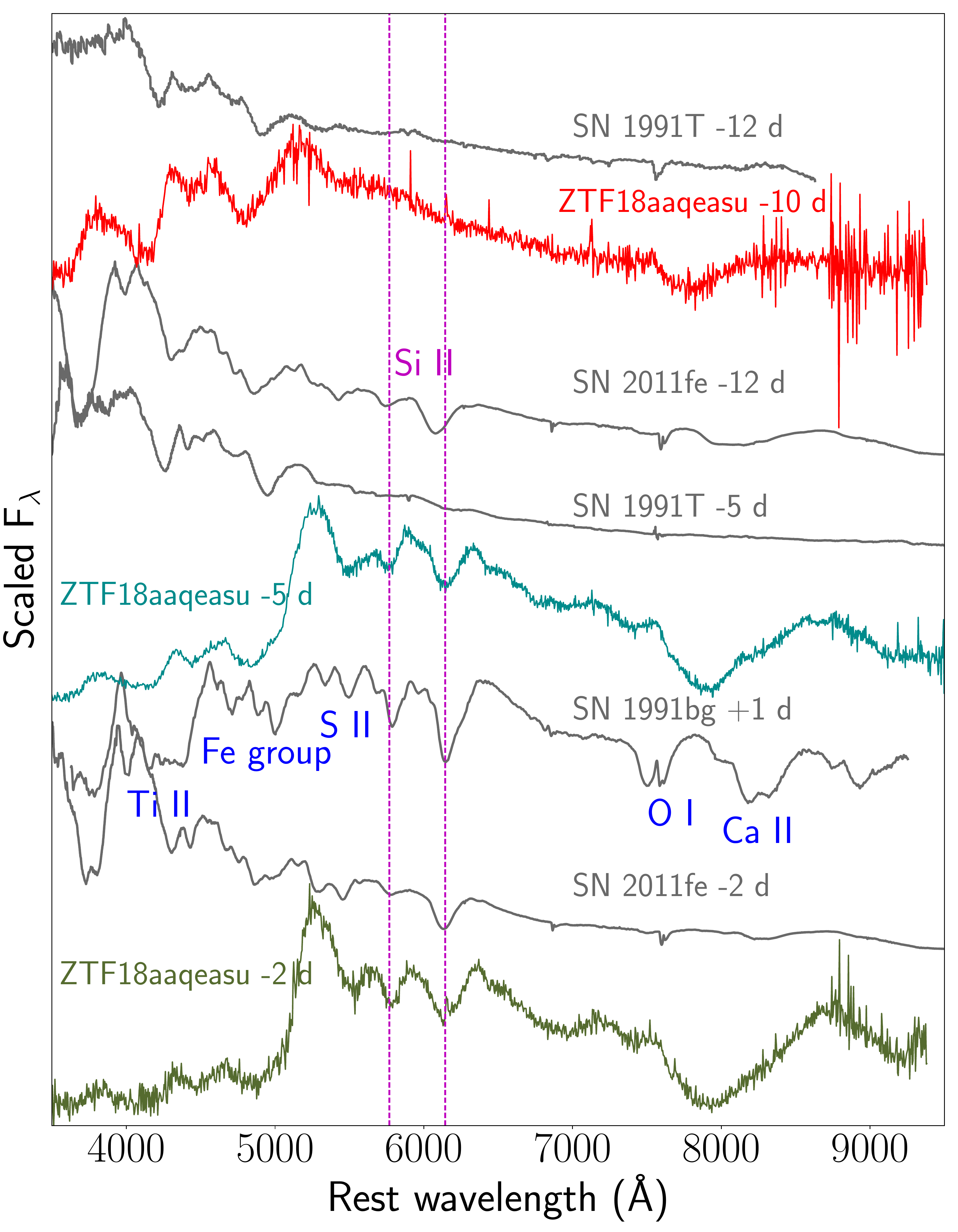}
\includegraphics[width=0.49\textwidth]{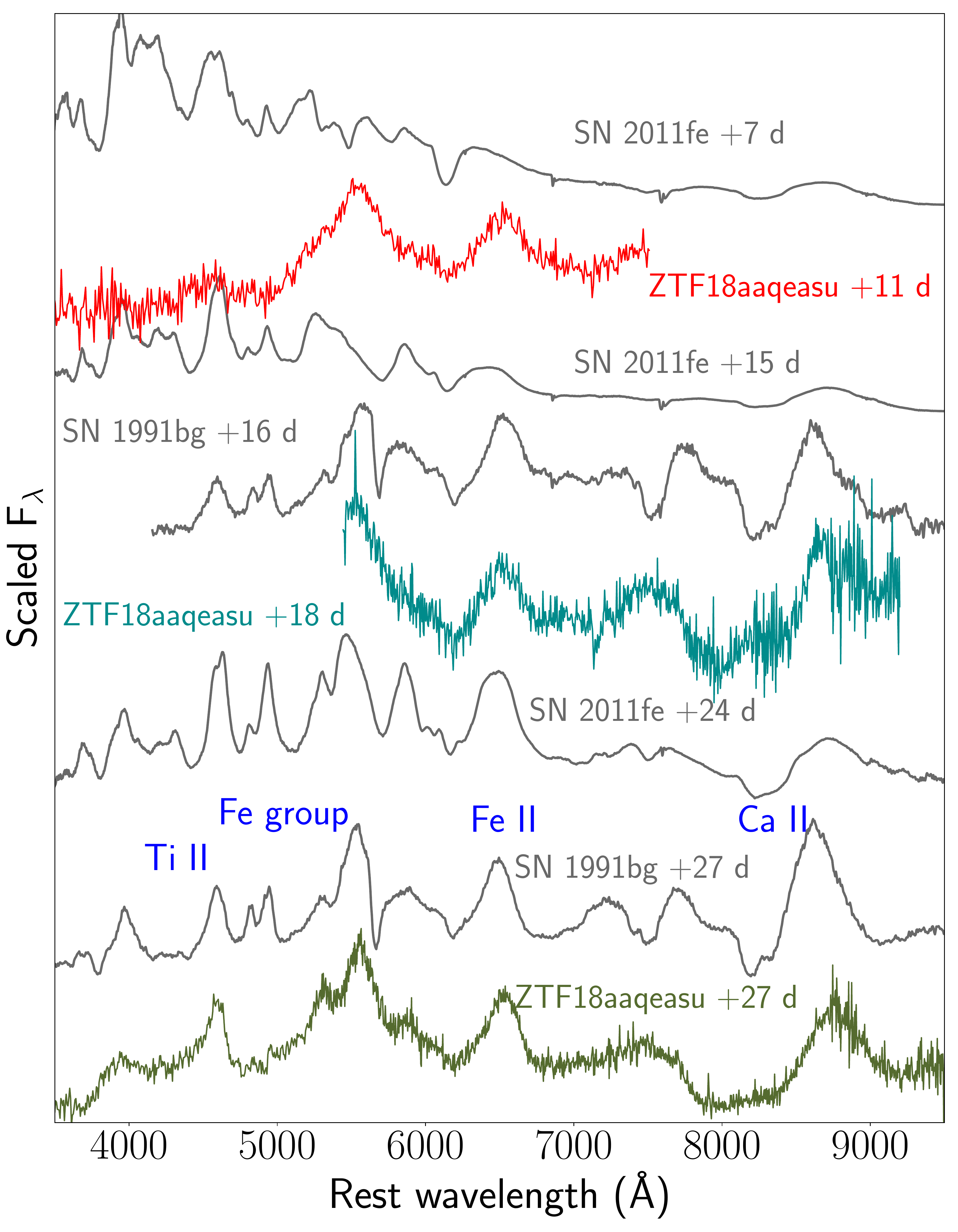}
\caption{Comparison of the pre-maximum (left) and post-maximum (right) spectra of ZTF18aaqeasu to the normal Type Ia SN 2011fe \citep{Maguire2014}, the over-luminous SN 1991T \citep{Filippenko1992a} and the sub-luminous SN 1991bg \citep{Filippenko1992b}. The magenta dashed lines in the left panel indicate Si II lines at a velocity of 10,000\,km\,s$^{-1}$. Prominent emission / absorption features are marked in both panels. Spectroscopic data for the comparison SNe were obtained from the WISEReP repository \citep{Yaron2012}.}
\label{fig:specComp}
\end{figure*}

We show a comparison of the pre-maximum and post-maximum spectroscopic evolution of ZTF\,18aaqeasu with other Type Ia SNe in Figure \ref{fig:specComp}. The earliest spectrum of the source was obtained $\approx 10$\,days before $r$-band peak and exhibits blue continua with broad absorption features blue-wards of $\approx 5500$\,\AA\,, notably without any Si II features that are characteristic of normal and sub-luminous Type Ia SNe at similar phases (Figure \ref{fig:specComp}). However, there are several similarities between this spectrum and SN 1991T at a similar phase, particularly in the absence of a Si II feature and the presence of broad absorption features of Ti II and Fe group elements near 4100\,\AA\, and 4700\,\AA\, respectively. This spectrum also shows signatures of an absorption feature in the Ca II NIR triplet.\\

Subsequent spectra taken at $\approx 5$ and $\approx 2$\,days before $r$-band peak exhibit the hallmark Si II absorption features found in Type Ia SNe near peak light. Using the minimum of the Si II P-Cygni profile, we measure a photospheric velocity of $\approx 10,500$\,km\,s$^{-1}$. However, the bluer parts of the spectra exhibit unusually strong line blanketing features leading to nearly complete absorption of flux blue-wards of 5000\,\AA. Comparing with the subluminous Type Ia SN 1991bg at a similar phase, we attribute this absorption to complete line blanketing by Fe group elements and Ti II. To our knowledge, such strong line blanketing features (and consequent red $g-r$ colors) have never been previously seen in any variant of a Type Ia SN at peak light. The Ca II triplet also develops into a deep, high velocity ($\approx 25,000$\,km\,s$^{-1}$) absorption feature at $7500$--$8500$\,\AA\, near peak.\\

Post-maximum spectra starting from $\approx 10$\,days after peak (Figure \ref{fig:specComp}) begin to develop broad emission features suggesting a transition to the optically thin phase. The only NIR spectrum taken at $+ 18$\,days exhibits a deep absorption feature at $\approx 9950$\,\AA. If associated with He I at 1.083\,$\mu$m, the corresponding absorption velocity would be $\approx 26,000$\,km\,s$^{-1}$. Spectra obtained at $\approx 27$ and $\approx 52$\,days after peak are similar to the Type Ia SN 1991bg at similar phases, and exhibit emission lines of Fe group elements, Ti II and Ca II. The late-time similarities and Fe-group dominated nucleosynthesis suggest a thermonuclear origin of the explosion, consistent with ZTF\,18aaqeasu representing an unusual variant of a Type Ia-like SN. However, the peculiar spectral features observed at peak light are unique to ZTF\,18aaqeasu and warrant further inspection with respect to a possible explosion mechanism.

\section{Model Comparisons}
\label{sec:models}

\begin{figure*}[!ht]
\centering
\includegraphics[width=0.49\textwidth]{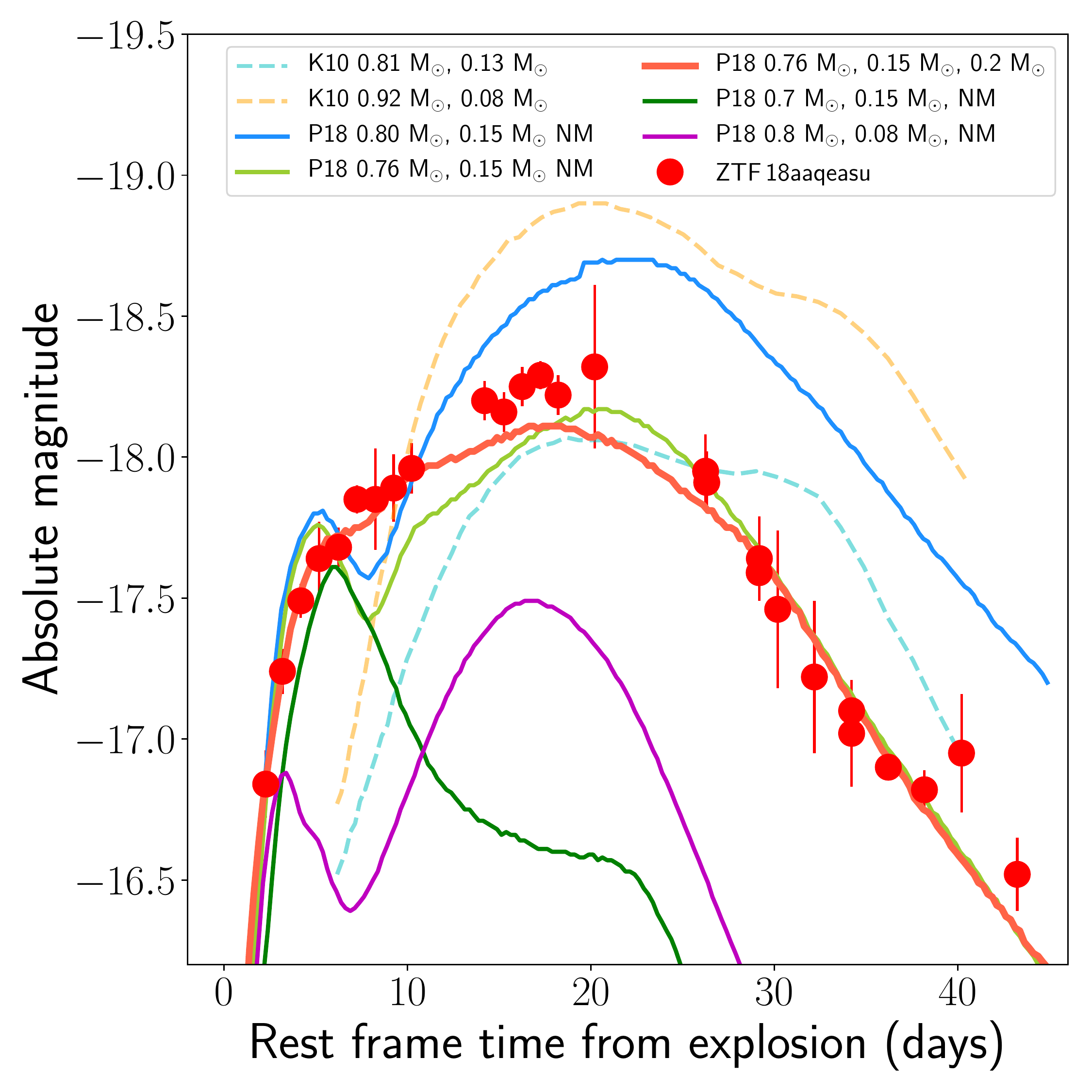}
\includegraphics[width=0.49\textwidth]{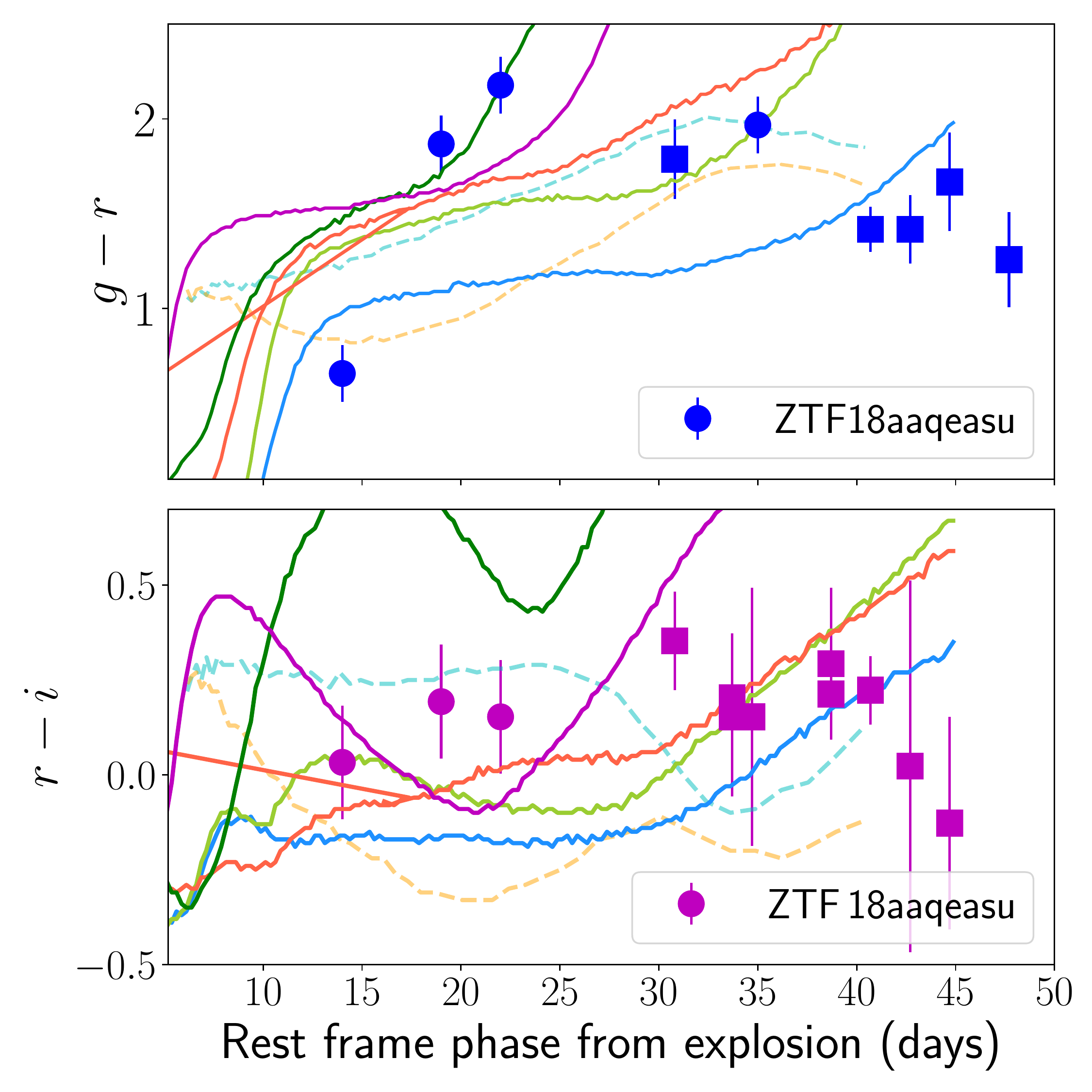}
\caption{Comparison of the photometric evolution of ZTF\,18aaqeasu with that of the He shell double detonation models in \citet{Kromer2010} (K10; dashed lines) and \citet{Polin2018} (P18; solid lines). The model parameters are indicated in the legend as (WD mass, shell mass, mixing length), and where NM stands for No Mixing. The left panel shows the $r$-band evolution while the right panels show the $g-r$ and $r-i$ color evolution with the same color schemes for the models. On the right panel, circles denote colors derived from the spectra while squares denote colors derived from photometry.}
\label{fig:model_phot}
\end{figure*}

\begin{figure*}[!ht]
\centering
\includegraphics[width=0.49\textwidth]{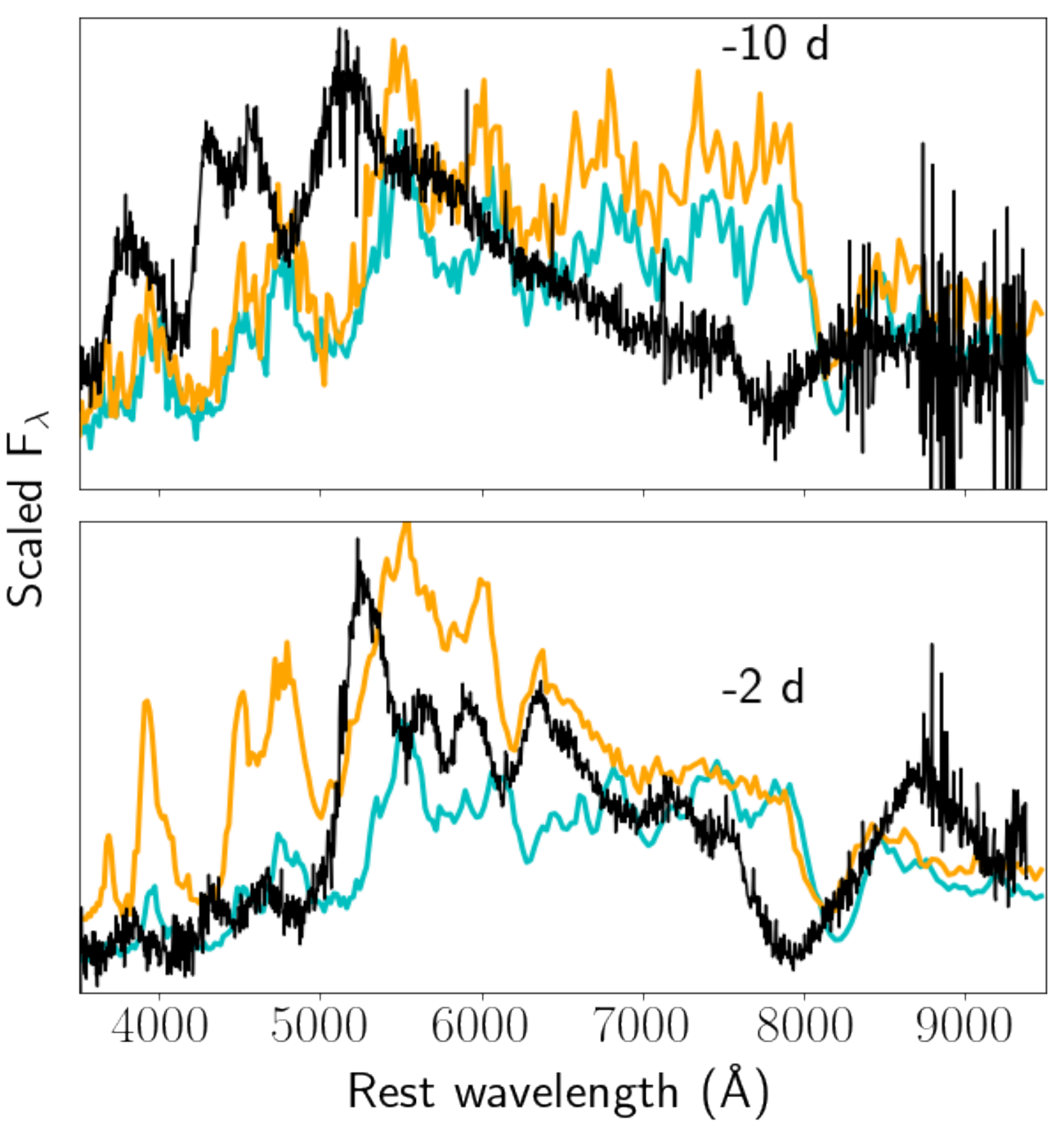}
\includegraphics[width=0.49\textwidth]{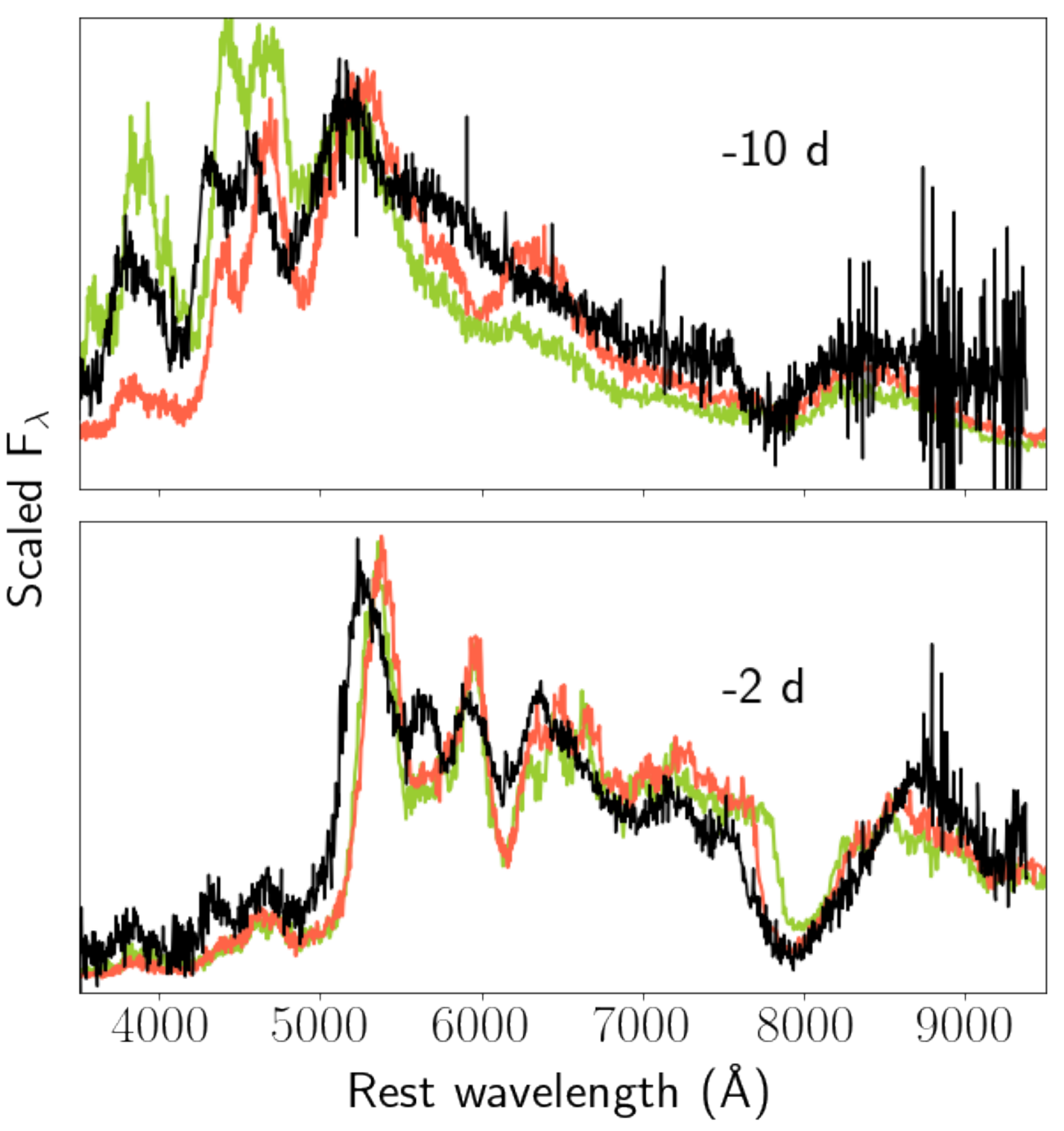}
\caption{Comparison of the spectra of ZTF\,18aaqeasu at and before peak light (shown as black lines) to models of He shell double detonations from \citet{Kromer2010} (left; for the two models shown in the light curve comparison) and \citet{Polin2018} (right; for two cases of ejecta mixing with the best fit WD and shell mass from the light curves). In both panels, the epochs of spectroscopic comparison are indicated according to the phase of the light curve (days from $r$-band peak), and the model spectra have the same color scheme as in Figure \ref{fig:model_phot}.}
\label{fig:model_spec}
\end{figure*}

The strong line blanketing features of Fe group elements observed in the peak spectra suggest that the outer layers of the ejecta are unusually rich in Fe group elements. The presence of such radioactive material in the outer ejecta would also be consistent with the fast rise observed in the early light curve \citep{Piro2014}, and is a hallmark signature of the decay of radioactive elements in the outermost ejecta (primarily $^{48}$Cr and $^{52}$Fe with half-lives of 0.90\,days and 0.35\,days respectively) for an explosion powered by a He shell detonation (e.g. \citealt{Nugent1997,Kromer2010,Polin2018}). This scenario would naturally explain the high velocity Ca II features observed near peak light, as a known He detonation product in the outer layers of the ejecta \citep{Fink2010,Kromer2010,Moore2013}.\\

We thus compare the properties of ZTF\,18aaqeasu to  simulations in the grid of models of He shell double detonations in \citet{Kromer2010}, who modeled the observable signatures for the minimum shell masses found in \citet{Bildsten2007}. We show a comparison of the $r$-band light curve, as well as the $g-r$ and $r-i$ colors of ZTF\,18aaqeasu to their angle averaged light curves in Figure \ref{fig:model_phot} (dashed lines). As shown, their lowest core mass model (with a WD core mass of 0.81\,\Msun; their Model 1) shows similarities to the data both in terms of absolute luminosity and red colors near peak, although the light curve shapes are not well matched. The larger luminosity of models involving higher mass WDs (yellow dashed lines) disfavor WD masses $\gtrsim 0.8$\,\Msun. \\

We also compare the light curves to the larger grid of shell and WD masses presented in the simulations of \citet{Polin2018}, including additional simulations performed to match the data (using the radiative transfer code SEDONA; \citealt{Kasen2006}), as solid lines in Figure \ref{fig:model_phot}. For the additional simulations, we model the explosion of a 50\% carbon + 50\% oxygen WD with an isentropic helium shell. The hydrodynamics and nuclear processes are modeled from the helium ignition until the ejecta reaches homologous expansion using the compressible hydrodynamics code Castro \citep{Almgren2010}. Once the explosion reached homology radiative transport calculations are performed using SEDONA to produce light curves and spectra from the ejecta. \\

Since the early rise in the light curve is powered by the radioactive He shell detonation products ($^{48}$Cr and $^{52}$Fe), we can constrain the mass of the shell. We note that a model involving a fixed WD mass of $0.8$\,\Msun and a shell of $0.15$\,\Msun shell reasonably reproduces the early rise, while the same WD with a smaller $0.08$\,\Msun shell under-predicts the $r$-band luminosity on the early rise. We then constrain the WD mass to be between $0.7$ and $0.8$\,\Msun by noting that models involving a $0.8$\,\Msun WD and $0.7$\,\Msun WD (each with a $0.15$\,\Msun shell) have a higher and lower peak luminosity than ZTF\,18aaqeasu respectively. With these constraints, we find that a model with a $0.76$\,\Msun WD and a $0.15$\,\Msun shell reproduces the overall $r$-band evolution. The corresponding synthesized $^{56}$Ni mass is $0.18$\,\Msun.\\

The light curves in these models also exhibit an early peak and decline, arising from the decay of radioactive material in the outer ejecta and the assumption of no mixing. Although we do observe signatures of a fast rising early peak, we do not have evidence of a decline as in the models, suggesting a possible influence of mixing in the ejecta. Hence, we also show an additional model with the best-fit WD mass and shell mass, but with ejecta mixed across a zone of $0.18$\,\Msun that is applied before performing the radiative transport. This model reproduces the early-time rise and the overall light curve. Although the red colors of the source near peak light are also well reproduced, the models become redder with time much faster than observed in the data. These discrepancies likely arise due to assumptions of local thermodynamic equilibrium (LTE) in SEDONA, which break down as the source transitions to the optically thin nebular phase at $\approx 20$\,days after peak light. \\
 
We also compare the spectroscopic properties of ZTF\,18aaqeasu to double detonation models in Figure \ref{fig:model_spec}. While there are several similarities between the model spectra of \citet{Kromer2010} and spectra near peak light ($\approx 2$\,days before peak light), including the strong line blanketing features below 5000\,\AA\ and deep Ca II absorption features, their limited grid of models does not correctly reproduce the line velocities and strengths of the prominent Si II and Ca II absorption features. Additionally, these models exhibit strong line blanketing features below 5000\,\AA\, and Si II absorption even at $\approx 10$\,days before peak, unlike the observed properties of this event where Si II lines appear only near peak light. \\

Comparing to the best-fit light curve model from \citet{Polin2018}, we find a better match to the observed spectra both before and at peak light (Figure \ref{fig:model_spec}). In the case of the early spectra ($\approx 10$\,days before peak), the models exhibit blue continua in the unmixed case similar to the data, although the absorption features become more prominent in the case of mixed ejecta as expected. The peak light spectra ($\approx 2$\,days before peak) are well reproduced both in terms of line velocities and strengths, and are not appreciably affected by mixing in the ejecta. Taken together, we find that a model involving the detonation of a $\approx 0.15$\,\Msun He shell on a $\approx 0.75$\,\Msun WD reproduces the observed signatures of the event after including mixing of $\approx 0.2$\,\Msun in the outer ejecta.

\section{Discussion and Conclusion}
\label{sec:discussion}

In this paper, we have presented observations of the transient ZTF\,18aaqeasu, and have shown it to be
a unique supernova. While its photometric and nucleosynthetic properties share several similarities to subluminous Type Ia SNe, its peak photospheric spectra are marked by extremely strong line blanketing features and red colors, unlike any previously observed Type Ia SN. By comparing the data to a grid of models, we show that the observed properties can be well explained by the detonation of a massive ($\approx 0.15$\,\Msun) He shell on a sub-Chandrasekhar mass ($\approx 0.75$\,\Msun) white dwarf. In particular, it is important to note that the He shell mass inferred is much larger than the thin He shells required to explain the properties of the broader population of Type Ia SNe ($\lesssim 0.01$\,\Msun; \citealt{Kromer2010,Sim2010,Shen2018,Polin2018}).\\

The inferred shell and core masses are consistent with what is predicted in the well studied He burning star donor scenario of a $0.48$\,\Msun sdB star losing matter to a C/O WD \citep{Nomoto1982a, Woosley1994, Woosley2011}. While much of the early work assumed a constant accretion rate, more recent calculations \citep{Brooks2015,Bauer2017} have self consistently calculated the evolution of the donor with mass loss and a self-consistent varying accretion rate. The properties inferred for ZTF\,18aaqeasu are remarkably close to the recent calculation of \citet{Bauer2017} for the evolution of the known sdB + WD binary CD $-30^\circ 11223$  \citep{Geier2013}, with \citet{Bauer2017} finding a He shell mass of $0.16$\,\Msun on the initial 
WD of mass $0.76$\,\Msun at the time of detonation. As this is the forward evolution of a known system, it’s possible to say that CD $-30^\circ 11223$ is indeed an example of the progenitor of ZTF\,18aaqeasu-like events.\\

The relatively luminous and slow evolving light curve of ZTF\,18aaqeasu (compared to e.g. SN 1991bg-like events) suggests that similar events should be easily detectable in a reasonably large volume of the local universe (out to $z \approx 0.1$ for a limiting magnitude of $r = 20.5$\,mag). However, no such event has been reported in previous studies of large samples of thermonuclear SNe (e.g., \citealt{Hicken2009,Maguire2014,Krisciunas2017,Scolnic2018}). To date, tentative evidence for only one other example of a relatively thin He shell detonation has been presented in \citet{Jiang2017} (see also \citealt{Polin2018}), although their modeling implies a more massive WD with a smaller He shell than is needed to explain ZTF\,18aaqeasu. Thus, ZTF\,18aaqeasu being the first of its kind suggests that massive He shell double detonations must be intrinsically rare in the population of thermonuclear SNe. \\

\section*{Acknowledgements}
This work was supported by the GROWTH project funded by the National Science Foundation under PIRE Grant No 1545949. This research benefited from interactions at a ZTF Theory Network meeting, funded by the Gordon and Betty Moore Foundation through Grant GBMF5076. ZTF is supported by the National Science Foundation and a collaboration including Caltech, IPAC, the Weizmann Institute for Science, the Oskar Klein Center at Stockholm University, the University of Maryland, the University of Washington, Deutsches Elektronen-Synchrotron and Humboldt University, Los Alamos National Laboratories, the TANGO Consortium of Taiwan, the University of Wisconsin at Milwaukee, and Lawrence Berkeley National Laboratories. Operations are conducted by COO, IPAC, and UW. Alert distribution is provided by DIRAC@UW \citep{Patterson2018}.\\

We thank Markus Kromer and Stuart Sim for providing their model data for comparisons. We thank A. Goobar, A. Gal-Yam, E. O. Ofek, J. Fuller, T. Kupfer and A. V. Filippenko for valuable discussions. Part of this research was carried out at the Jet Propulsion Laboratory, California Institute of Technology, under a contract with the National Aeronautics and Space Administration. Some of the data presented herein were obtained at the W.M. Keck Observatory, which is operated as a scientific partnership among the California
Institute of Technology, the University of California and the National Aeronautics and Space Administration. The Observatory was made possible by the generous financial support of the W.M. Keck Foundation. The authors wish to recognize and acknowledge the very significant cultural role and reverence that the summit of Mauna Kea has always had within the indigenous Hawaiian community. We are most fortunate to have the opportunity to conduct observations from this mountain.  These results made use of the Discovery Channel Telescope at Lowell Observatory. Lowell is a private, non-profit institution dedicated to astrophysical research and public appreciation of astronomy and operates the DCT in partnership with Boston University, the University of Maryland, the University of Toledo, Northern Arizona University and Yale University. The upgrade of the DeVeny optical spectrograph has been funded by a generous grant from John and Ginger Giovale.

\bibliographystyle{aasjournal}
\bibliography{ZTF18aaqeasu}

\end{document}